\documentclass[pre,twocolumn,showpacs]{revtex4}   \newcount\tipo\tipo=1 

 \let\g=\gamma \let\d=\delta
 \let\z=\zeta 
 \let\l=\lambda \let\m=\mu 
   \let\s=\sigma \let\t=\tau
 \let\f=\varphi  
  \let\D=\Delta 
 \let\P=\Pi  \let\F=\Phi
 \let\O=\Omega 
\let\dpr=\partial\let\0=\noindent\let\fra=\frac
\def\media#1{\langle{#1}\rangle}
\def\*{\vskip3mm}

\def\Onlinecite#1{[\onlinecite{#1}]}
\def\tende#1{\ \vtop{\ialign{##\crcr\rightarrowfill\crcr
\noalign{\kern-1pt\nointerlineskip} \hglue3.pt${\scriptstyle%
#1}$\hglue3.pt\crcr}}\,}
\newcommand\defi{\,{\buildrel def \over =}\,}
\newcommand\revtex{{R\kern-1mm\lower0.5mm\hbox{E}\kern-0.6mm V\kern-0.5mm%
\lower0.5mm\hbox{T}\kern-0.5mm E\kern-.5mm \lower0.5mm\hbox{X}}}

\begin{document}
\preprint{FM 04-04} 

\title{Fluctuation theorem revisited}
\author{Giovanni Gallavotti}
\affiliation{Rutgers Hill Center, I.N.F.N. Roma1, ENS Paris}
\relax 
\maketitle
\relax 

\kern-4cm
\0\date{\today}

\kern1cm

\0{\bf Abstract}: Recently the ``Fluctuation theorem'' has been
criticized and incorrect incorrect contents have been atributed to it.
Here I reestablish and comment the original statements.

\vskip3mm

\relax 

\0{\bf Fluctuations}
\vskip3mm

Mathematically the {\it Fluctuation theorem} is a property of the
phase space contraction of an Anosov map $S$, called {\it time
evolution}, which is {\it time reversible}. The possible connection
between the fluctuation theorem and Physics is a different matter that
I will not discuss here: there are many places where this is done in
full detail, \Onlinecite{Ru99,Ga99,Ga00}.

I shall denote by $\O$ the phase space (a smooth finite boundaryless
manifold) and by $\s(x)$ the phase space contraction

$$\s(x)=-\log |\det \dpr_xS(x)|$$
Time reversal will be an {\it isometry} of phase space $I$ such that

$$I S = S^{-1}I, \qquad \s(Ix)=-\s(x)$$ 
It has been shown that there exists a unique probability distribution,
called the {\it statistics of the motion} or the {\it SRB
distribution}, $\m$ such that for all points $x\in\O$, excepted those
in a set of $0$ volume, it is

$$\lim_{\t\to\infty} \fra1\t \sum_{t=0}^{\t-1} F(S^tx)\defi
\media{F}=\int_\O F(y)\m(dy)$$
for all smooth observables $F$ defined on phase space.

It is intuitive that ``phase space cannot expand''; this is expressed
by the following result of Ruelle, \Onlinecite{Ru96},

\*
{\it Proposition: If $\s_+\defi\media{\s}$ it is $\s_+\ge0$} 
\*

Clearly if $S$ is volume preserving it is $\s_+=0$. If $\s_+>0$ the
system does not admit any stationary distribution which is absolutely
continuous with respect to the volume.

This motivates calling systems for which $\media{\s}>0$ ``dissipative''
and calling volume preserving systems ``conservative''.

For Anosov systems which are ``transitive'' ({\it i.e.} with a dense orbit),
reversible and dissipative one can define the dimensionless
phase space contraction, a quantity related to entropy creation rate
(see \Onlinecite{Ga03b}) averaged over a time interval of size $\t$.
This is

$$p= \fra1{\s_+ \t }\sum_{-\t/2}^{\t/2-1} \s(S^k x)$$
provided {\it of course} $\s_+>0$. 

Then for such systems the probability with respect to the stationary
state, {\it i.e.} to the SRB distribution $\m$, that the variable $p$ takes
values in $\D=[p,p+\d p]$ can be written as $\P_\t(\D)=e^{\t
\max_{p\in \D}\z(p) + O(1)}$ where $\z(p)$ is a suitable function and
$O(1)$ refers to the $\t$--dependence at fixed $p,\d p$ for all
intervals $\D$ contained in an open interval $(p^*_1,p^*_2)$ (this is
often expressed as $\lim_{\t\to\infty}\fra1\t \log \P_\t(p)=\z(p)$ for $\
p^*_1<p<p^*_2$). The function $\z(p)$ would be called in probability
theory the {\it rate function} for the large deviations of $\s(x)$.

The function $\z(p)$ is analytic in $p$ in the interval of
definition $(p^*_1,p^*_2)$ and {\it convex}.  In fact more is true and
one can prove the following {\it fluctuation theorem}:

\*
{\it Proposition: In transitive
time reversible Anosov systems the rate function $\z(p)$ for the phase space
contraction $\s(x)$ is analytic and strictly convex
in an interval $(-p^*,p^*)$ with $+\infty>p^*\ge1$ and $\z(p)=-\infty$ for
$|p|>p^*$.
Furthermore 

$$\z(-p)=\z(p)-p\s_+, \qquad {\rm for} \qquad |p|<p^* $$
which is called the ``fluctuation relation''.
}
\*

The result is adapted from a theorem by Sinai who proves analyticity
and convexity. Strict convexity follows from a theorem of Griffiths
and Ruelle which shows that the only way strict convexity could fail
is if $\s(x)=\f(Sx)-\f(x)+ c$ where $\f(x)$ is a smooth function
(typically a Lipschitz continuous function) and $c$ is a constant, see
propositions (6.4.2) and (6.4.3) in \Onlinecite{GBG04}. The constant
vanishes if time reversal holds and $\s(x)=\f(Sx)-\f(x)$ contradicts
the assumption that $\s_+>0$ (because $\t^{-1}\sum_{-\t/2}^{\t/2-1}
\s(S^k x)=\t^{-1}(\f(S^{\t/2-1}x)-\f(S^{-\t/2}x))\to 0$ as $\t\to\infty$.
The value of $p^*$ must be $p^*\ge1$ otherwise $p^*<1$ and the average
of $p$ could not be $1$ (as it is by its very definition). The
fluctuation relation is in \Onlinecite{GC95} and is properly called
the {\it fluctuation theorem} (a name later given to other very
different relations with remarkable confusion, \Onlinecite{CG99}).
The theorem can be extended to Anosov flows ({\it i.e.} to systems evolving
in continuous time), \cite{Ge98}.
\*

\0{\it Remarks:} The relation was discovered in a numerical
experiment, \Onlinecite{ECM93}, and proved in \Onlinecite{GC95}.  For
finite $\t$ the function $\z(p)$ and $\z(-p)$ are replaced by
$\z_\t(p),\z_\t(-p)$ which differ from their limits as $\t\to\infty$ by a
quantity bounded by a constant uniforlmly in any closed interval of
$(-p^*,p^*)$ \*

Sometimes one does not consider the above $p$ but a quantity
$a=\t^{-1}\sum_{j=-\t/2}^{\t/2-1} \s(S^j x)$ and the result can be
written

$$\widetilde\z(-a)=\widetilde\z(a)-a,\qquad {\rm for}\ |a|< p^*\s_+$$
where $\widetilde\z(a)$ is trivially related to $\z(p)$.  This form
dangerously suggests that in the case of systems with $\s_+=0$ the
distribution of the variable $a$ is asymmetric (because the extra
condition $|a|<p^*\s_+$ might be forgotten, see \Onlinecite{Ga04}).

Note that $p^*$ is certainly $<+\infty$ because the variable $\s(x)$ is
bounded (being continuous on phase space, {\it i.e.}  on the bounded manifold
on which the Anosov map is defined). 

However no confusion should be made between $p^*\s_+$ and
$\sigma_{max}\defi\max |\s(x)|$: unlike $\sigma_{max}$ the quantity
$p^*$ is a {\it non trivial} dynamical quantity, independent on the
metric used on phase space to measure distances, hence volume. It is
very easy to build examples in which $p^*\s_+< \sigma_{max}$ and in
fact the ratio between the r.h.s and the l.h.s two quantities can be
even infinite ({\it e.g.} in conservative cases when $p^*\s_+=0$ but
$\s(x)$ is not identically $0$ because of the metric used.

Note that even in the conservative cases we can define on phase space
a time reversal invariant metric, {\it i.e.} such that $I$ is an
isometry, whose volume elements do not verify Liouville's theorem,
see appendix). This point has not been always understood and the
confusion has in fact been made, at least once, in the published
literature with nefast consequences.
\*

{\it The fact is that $p^*$ and $\max |a|$ are dynamically determined,
non trivial, quantities and one cannot ``assume'' their value, see comment
13 in reference \cite{Ga04}.}
\*

Considering more closely the cases $\s_+=0$ it follows that
$\s(x)=\f(Sx)-\f(x)+c$ again by the mentioned result of Griffiths and
Ruelle (essentially the same mentioned above) and $c=0$ by time
reversal. Hence the variable

$$a=\fra1\t\sum_{j=-\t/2}^{\t/2-1} \s(S^jx)$$
is {\it bounded} and tends {\it uniformly} to $0$. One could repeat
the theory developed for $p$ when $\s_+>0$ but one would reach the
conclusion that $\widetilde \z(a)=-\infty$ for $|a|>0$ and we see that
the result is trivial. In fact in this case it follows that that the
system admits an absolutely continuous SRB distribution. The
distribution of $a$ is symmetric (trivially by time reversal
symmetry) and becomes a delta function around $0$ as $t\to\infty$.

Nevertheless the fluctuation relation is {\it non trivial} in cases in
which the map $S$ depends on parameters ${\bf E}=(E_1,\ldots,E_n)$ and
becomes volume preserving (``conservative'') as ${\bf E}\to0$: in this
case $\s_+\to0$ as ${\bf E}\to{\bf0}$ and one has to rewrite the fluctuation
relation in an appropriate way to take a meaningful limit. 

The result is that the limit as ${\bf E}\to0$ of the fluctuation relation
in which both sides are divided by ${\bf E}^2$ makes sense and yields (in
the case considered here of transitive Anosov dynamical systems)
relations which are non trivial and that can be interpreted as giving
Green--Kubo formulae and Onsager reciprocity for transport
coefficients, \Onlinecite{Ga96a}.

In fact the very definition of the duality between currents and fluxes
so familiar in nonequilibrium thermodynamics since Onsager can be set
up in such systems using as generating function the $\s_+$ regraded as
a function of ${\bf E}$. Note that the fluxes are usually ``currents''
divided by the temperature: therefore via the above interpretation one
can try to define the temperature even in nonequilibrium situations,
\Onlinecite{GC04}.  
\* 

\0{\bf A test} 
\*

Although a check of the fluctuation relation is difficult nevertheless
it has been performed in several cases. Little attention has been
dedicated, however, until recently to one rather striking prediction
valid under suitable {\it further assumptions} which are proposed in
\Onlinecite{BG97} where they are called {\it axiom C} and {\it pairing
rule} and are presented as possibly quite general the first, and as at
least approximately valid in several examples the second. The
prediction of the first assunption is that if under strong forcing (or
strong dissipation) the attracting set ({\it i.e.} the closure af the
attractor) for the dynamics becomes slim and occupies a region of
dimensionality lower than that of phase space, then the a new symmetry
with the same properties but which leaves the attracting set invariant
is generated: ``time reversal is unbreakable'',
\Onlinecite{BG97,Ga98}. Hence the area contraction on the attracting
set will verify the fluctuation relation.

In general,however, the area elements on the attracting set will be
very difficult to measure: but at least in the cases in which the
second assumption holds the prediction is that their contraction is
related to the total volume contraction $\s(x)$ on the full phase
space which is much easier to access. Furthermore the fluctuation
relation will hold in the form $\z(p)=\z(-p)-p\,\g\,\s_+$ {\it with
$\g$ equal to the ratio of the dimension of the attracting set to the
dimension of the full phase space}. This in particular implies $\g
<1$: which is a surprising and apparently counterintuitive result: a
naive view of the attraction mechanism: {\it i.e.} a uniform contraction of
the transversal directions would in fact lead to $c>1$.

The reason why one expects $\g >1$ if the contraction transversal to the
attracting set is constant and equal to $\l_0$ (a property which,
however, is incompatible with the pairing rule) is the following. The
total contraction could be written $\s(x)=\s_0(x)+\l_0$ then, setting
$c=\fra{\overline\s_0}{\overline\s_0+\l_0}$ with $\overline\s_0=\s_{0+}$, and
$\l=\fra{\l_0}{\overline\s_0+\l_0}$, it is $p= c\, p_0+\l$ and

$$\z(p)-\z(-p)=\z_0(\fra{p-\l}c)-\z_0(\fra{-p-\l}c)$$
Developing the expression in powers of $\l$ and using the fluctuation
relation for $p_0$ and assuming a gaussian distribution for $p_0$
(which is related to the validity of the Green--Kubo relations, see
\cite{ZRA04}) we get $\z(p)-\z(-p)=(1+\l)\,\s_+\, p$, {\it i.e.} $\g=1+\l>1$
for small $\l$ and small $\s_{0+}$.

The results in \cite{ZRA04} might prelude to the first check of the quite
striking prediction that the coefficient $\g$ is $<1$ because in this
case the pairing rule seems to be verified within a good
approximation (although not exact,
\Onlinecite{BCP98}).  
\vskip3mm

\0{\bf Appendix: 
\it No relation between $p^*$, $\max |a|$  and $\s_{max}=\max|\s(x)|$:
some explicit counterexamples} 
\vskip3mm

The simplest example (out of many) is provided by the simplest
conservative system which is strictly an Anosov transitive system and
which has therefore an SRB distribution: this is the geodesic flow on
a surface of constant negative curvature, \Onlinecite{BGM98}.  I
discuss here an evolution in continuous time because the matter is
considered in the literature for such systems, \Onlinecite{Ga04} (even
simpler examples are possible for time evolution maps).

The phase space $M$ is compact, time reversal is just momentum
reversal and the natural metric, induced by the Lobatchesky metric
$g_{ij}(q)$ on the surface, is time reversal invariant: the SRB
distribution is the Liouville distribution and $\s(x)\equiv0$. However
one can introduce a function $\F(x)$ on $M$ which is very large in a
small vicinity of a point $x_0$, arbitrarily selected, constant
outside a slightly larger vicinity of $x_0$ and positive everywhere. A
new metric could be defined as $g_{new}(x)=(\F(x)+\F(Ix)) g(x)\defi
F(x) g(x)$: it is still time reversal invariant but its volume
elements {\it will no longer be invariant} under time evolution. The
rate of change of phase space volume in the new metric will be
$\sigma_{new}(x)={\cal L}F(x)$ where ${\cal L}$ is the Liouville
operator. Since $\F$ is arbitrary one can achieve a value of
$\s_{new}(x)$ as large as wished by fixing suitably the function
$\F$. 

Nevertheless $a=\fra1\t \int_0^\t \s_{new}(S_t x) dt= \t^{-1}(F(S_\t
x)-F(x))\tende{\t\to\infty}0$. This contradicts statements existing in
the literature, \cite{Ga04}, which claim that in such case the
quantity $a$ will verify the relation
$\widetilde\z(a)=\widetilde\z(-a)- a$: in fact the distribution of $a$
will be a delta function at $0$ hence the relation cannot hold with
$\widetilde\z(a)$ finite for $a\ne0$, even though some go as far as
claiming that such erroneous conclusion would follow by ``repeating'',
with minor adaptations, Ruelle's proof of the fluctuation theorem
(which would therefore be incorrect, which is not), see comments 12,13
and 23 in reference \cite{Ga04}.

It has been argued that even if $\s_+>0$ but small ({\it i.e.} the system is
close to equilibrium) and the system is Anosov the fluctuation
relation will not apply under certain thermostat mechanisms: but the
work \cite{BGM98} provides a counterxample (out of many others
possible) even to this statement, see \cite{Ga04}.  
\*

\kern1mm \0{\it Acknowledgements: I am indebted to
E.G.D. Cohen for many enlightening discussions.}


\begin{thebibliography}{19}
\expandafter\ifx\csname natexlab\endcsname\relax\def\natexlab#1{#1}\fi
\expandafter\ifx\csname bibnamefont\endcsname\relax
  \def\bibnamefont#1{#1}\fi
\expandafter\ifx\csname bibfnamefont\endcsname\relax
  \def\bibfnamefont#1{#1}\fi
\expandafter\ifx\csname citenamefont\endcsname\relax
  \def\citenamefont#1{#1}\fi
\expandafter\ifx\csname url\endcsname\relax
  \def\url#1{\texttt{#1}}\fi
\expandafter\ifx\csname urlprefix\endcsname\relax\def\urlprefix{URL }\fi
\providecommand{\bibinfo}[2]{#2}
\providecommand{\eprint}[2][]{\url{#2}}

\bibitem[{\citenamefont{Ruelle}(1999)}]{Ru99}
\bibinfo{author}{\bibfnamefont{D.}~\bibnamefont{Ruelle}},
  \bibinfo{journal}{Journal of Statistical Physics}
  \textbf{\bibinfo{volume}{95}}, \bibinfo{pages}{393} (\bibinfo{year}{1999}).

\bibitem[{\citenamefont{G.Gallavotti}(1999)}]{Ga99}
\bibinfo{author}{\bibnamefont{G.Gallavotti}}, \bibinfo{journal}{Open Systems
  and Information Dynamics} \textbf{\bibinfo{volume}{6}}, \bibinfo{pages}{101}
  (\bibinfo{year}{1999}).

\bibitem[{\citenamefont{Gallavotti}(2000)}]{Ga00}
\bibinfo{author}{\bibfnamefont{G.}~\bibnamefont{Gallavotti}},
  \emph{\bibinfo{title}{Statistical Mechanics. A short treati\-se}}
  (\bibinfo{publisher}{Springer Verlag}, \bibinfo{address}{Berlin},
  \bibinfo{year}{2000}).

\bibitem[{\citenamefont{Ruelle}(1996)}]{Ru96}
\bibinfo{author}{\bibfnamefont{D.}~\bibnamefont{Ruelle}},
  \bibinfo{journal}{Journal of Statistical Physics}
  \textbf{\bibinfo{volume}{85}}, \bibinfo{pages}{1} (\bibinfo{year}{1996}).

\bibitem[{\citenamefont{Gallavotti}(2003)}]{Ga03b}
\bibinfo{author}{\bibfnamefont{G.}~\bibnamefont{Gallavotti}},
  \bibinfo{journal}{cond-mat/0312657}  (\bibinfo{year}{2003}).

\bibitem[{\citenamefont{Gallavotti et~al.}(2004)\citenamefont{Gallavotti,
  Bonetto, and Gentile}}]{GBG04}
\bibinfo{author}{\bibfnamefont{G.}~\bibnamefont{Gallavotti}},
  \bibinfo{author}{\bibfnamefont{F.}~\bibnamefont{Bonetto}}, \bibnamefont{and}
  \bibinfo{author}{\bibfnamefont{G.}~\bibnamefont{Gentile}},
  \emph{\bibinfo{title}{Aspects of the ergodic, qualitative and statistical
  theory of motion}} (\bibinfo{publisher}{Springer Verlag},
  \bibinfo{address}{Berlin}, \bibinfo{year}{2004}).

\bibitem[{\citenamefont{Gallavotti and Cohen}(1995)}]{GC95}
\bibinfo{author}{\bibfnamefont{G.}~\bibnamefont{Gallavotti}} \bibnamefont{and}
  \bibinfo{author}{\bibfnamefont{E.}~\bibnamefont{Cohen}},
  \bibinfo{journal}{Physical Review Letters} \textbf{\bibinfo{volume}{74}},
  \bibinfo{pages}{2694} (\bibinfo{year}{1995}).

\bibitem[{\citenamefont{Cohen and G.Gallavotti}(1999)}]{CG99}
\bibinfo{author}{\bibfnamefont{E.}~\bibnamefont{Cohen}} \bibnamefont{and}
  \bibinfo{author}{\bibnamefont{G.Gallavotti}}, \bibinfo{journal}{Journal of
  Statistical Physics} \textbf{\bibinfo{volume}{96}}, \bibinfo{pages}{1343}
  (\bibinfo{year}{1999}).

\bibitem[{\citenamefont{Gentile}(1998)}]{Ge98}
\bibinfo{author}{\bibfnamefont{G.}~\bibnamefont{Gentile}},
  \bibinfo{journal}{Forum Mathematicum} \textbf{\bibinfo{volume}{10}},
  \bibinfo{pages}{89} (\bibinfo{year}{1998}).

\bibitem[{\citenamefont{Evans et~al.}(1993)\citenamefont{Evans, Cohen, and
  Morriss}}]{ECM93}
\bibinfo{author}{\bibfnamefont{D.}~\bibnamefont{Evans}},
  \bibinfo{author}{\bibfnamefont{E.}~\bibnamefont{Cohen}}, \bibnamefont{and}
  \bibinfo{author}{\bibfnamefont{G.}~\bibnamefont{Morriss}},
  \bibinfo{journal}{Physical Review Letters} \textbf{\bibinfo{volume}{70}},
  \bibinfo{pages}{2401} (\bibinfo{year}{1993}).

\bibitem[{\citenamefont{Gallavotti}(2004)}]{Ga04}
\bibinfo{author}{\bibfnamefont{G.}~\bibnamefont{Gallavotti}},
  \bibinfo{journal}{cond-mat 0402676}  (\bibinfo{year}{2004}).

\bibitem[{\citenamefont{Gallavotti}(1996{\natexlab{a}})}]{Ga96a}
\bibinfo{author}{\bibfnamefont{G.}~\bibnamefont{Gallavotti}},
  \bibinfo{journal}{Physical Review Letters} \textbf{\bibinfo{volume}{77}},
  \bibinfo{pages}{4334} (\bibinfo{year}{1996}{\natexlab{a}}).

\bibitem[{\citenamefont{Gallavotti and Cohen}(2004)}]{GC04}
\bibinfo{author}{\bibfnamefont{G.}~\bibnamefont{Gallavotti}} \bibnamefont{and}
  \bibinfo{author}{\bibfnamefont{E.}~\bibnamefont{Cohen}},
  \bibinfo{journal}{Physical Review E} \textbf{\bibinfo{volume}{69}},
  \bibinfo{pages}{035104 (+4)} (\bibinfo{year}{2004}).

\bibitem[{\citenamefont{Bonetto and Gallavotti}(1997)}]{BG97}
\bibinfo{author}{\bibfnamefont{F.}~\bibnamefont{Bonetto}} \bibnamefont{and}
  \bibinfo{author}{\bibfnamefont{G.}~\bibnamefont{Gallavotti}},
  \bibinfo{journal}{Communications in Mathematical Physics}
  \textbf{\bibinfo{volume}{189}}, \bibinfo{pages}{263} (\bibinfo{year}{1997}).

\bibitem[{\citenamefont{Gallavotti}(1998)}]{Ga98}
\bibinfo{author}{\bibfnamefont{G.}~\bibnamefont{Gallavotti}},
  \bibinfo{journal}{Physica D} \textbf{\bibinfo{volume}{112}},
  \bibinfo{pages}{250} (\bibinfo{year}{1998}).

\bibitem[{\citenamefont{Zamponi et~al.}(2004)\citenamefont{Zamponi, Ruocco, and
  An\-gelani}}]{ZRA04}
\bibinfo{author}{\bibfnamefont{F.}~\bibnamefont{Zamponi}},
  \bibinfo{author}{\bibfnamefont{G.}~\bibnamefont{Ruocco}}, \bibnamefont{and}
  \bibinfo{author}{\bibfnamefont{L.}~\bibnamefont{An\-gelani}},
  \bibinfo{journal}{cond-mat/0403579}  (\bibinfo{year}{2004}).

\bibitem[{\citenamefont{Bonetto et~al.}(1998)\citenamefont{Bonetto, Cohen, and
  Pugh}}]{BCP98}
\bibinfo{author}{\bibfnamefont{F.}~\bibnamefont{Bonetto}},
  \bibinfo{author}{\bibfnamefont{E.}~\bibnamefont{Cohen}}, \bibnamefont{and}
  \bibinfo{author}{\bibfnamefont{C.}~\bibnamefont{Pugh}},
  \bibinfo{journal}{Journal of Statistical Physics}
  \textbf{\bibinfo{volume}{92}}, \bibinfo{pages}{587} (\bibinfo{year}{1998}).

\bibitem[{\citenamefont{Bonetto et~al.}(2000)\citenamefont{Bonetto, Gentile,
  and Mastropietro}}]{BGM98}
\bibinfo{author}{\bibfnamefont{F.}~\bibnamefont{Bonetto}},
  \bibinfo{author}{\bibfnamefont{G.}~\bibnamefont{Gentile}}, \bibnamefont{and}
  \bibinfo{author}{\bibfnamefont{V.}~\bibnamefont{Mastropietro}},
  \bibinfo{journal}{Ergodic Theory and Dynamical Systems}
  \textbf{\bibinfo{volume}{20}}, \bibinfo{pages}{681} (\bibinfo{year}{2000}).

\bibitem[{\citenamefont{Gallavotti}(1996{\natexlab{b}})}]{Ga96}
\bibinfo{author}{\bibfnamefont{G.}~\bibnamefont{Gallavotti}},
  \bibinfo{journal}{Journal of Statistical Physics}
  \textbf{\bibinfo{volume}{84}}, \bibinfo{pages}{899}
  (\bibinfo{year}{1996}{\natexlab{b}}).

\end{thebibliography}

\bibliographystyle{apsrev} 
\*

\def\revtexz{{\bf
R\lower1mm\hbox{E}V\lower1mm\hbox{T}E\lower1mm\hbox{X}}} 

\0e-mail: {\tt giovanni.gallavotti@roma1.infn.it}\\
web: {\tt http://ipparco.roma1.infn.it}\\
Dip. Fisica, U. Roma 1, 00185, Roma, Italia\\

\end{document}